\documentstyle [prb,aps,epsf]{revtex}
\begin{document}
\widetext
\draft

\title{
A strongly correlated electron model for\\
the layered organic superconductors
$\kappa$-(BEDT-TTF)$_2$X  
}

\author{Ross H. McKenzie\cite{email}}

\address{School of Physics, University of New
South Wales, Sydney 2052, Australia}
                       
\date{Received March 17, 1998}

\author{Published as {\it Comments Cond. Matt. Phys.} {\bf 18,} 309 (1998)}

\maketitle

\begin{abstract}
The fascinating electronic properties of
the family of layered organic molecular crystals 
$\kappa$-(BEDT-TTF)$_2$X  
where X is an anion (e.g., X=I$_3$, Cu[N(CN)$_2$]Br, Cu(SCN)$_2$)
are reviewed.  These materials are particularly interesting because 
of similarities to
the high-$T_c$ cuprate superconductors 
including unconventional metallic properties and
competition between antiferromagnetism and
superconductivity.
The temperature dependence of electrical transport,
optical, and nuclear magnetic resonance
properties deviate significantly from those of
a conventional metal.
In particular, there appears to
be an effective Fermi energy of the order of 100 K
which is an order of magnitude smaller than 
predicted by band structure calculations.
The results of quantum chemistry calculations suggest          
that a minimal theoretical model that can describe
these materials is a Hubbard model on an
anisotropic triangular lattice
with one electron per site.
Appropriate parameter values for the model imply
that the electronic correlations are strong,
 significant magnetic frustration is present,
and the system is close to a metal-insulator transition.
Insight into the physics of this model can be obtained from recent
studies of the Hubbard model using a dynamical
mean-field approximation.
They are consistent with a low effective Fermi energy
and the unconventional temperature dependence of
many of the properties of
 the metallic phase.
Future directions
 are     suggested for both theoretical and experimental studies.
\\ 
\\
\end{abstract}

The discovery of heavy fermion metals,
high-temperature superconductivity
in copper oxides,
and colossal magnetoresistance in manganates
has stimulated extensive experimental
studies of these materials and
extensive theoretical studies 
of strongly correlated electron models.\cite{fulde}
A key feature of the copper oxides (cuprates)
is their layered or quasi-two-dimensional
structure which enhances the effects of
the interactions between the electrons
and leads to metallic properties
that are quite distinct from conventional metals.\cite{anderson,allen}
Organic conductors based on the 
BEDT-TTF [bis-(ethylenedithia-tetrathiafulvalene)]
 molecule have been widely studied because
prior to the discovery of fullerene superconductors
they had the highest transition temperature
($T_c \sim$ 10 K) of an organic superconductor.
 The family $\kappa-$(BEDT-TTF)$_2$X
where X is an anion (X=Cu[N(CN)$_2$]Br, for example)
are particularly interesting
because of similarities to the cuprates.\cite{mck,greene}
This article gives a brief review of these materials
focussing on how the properties of the metallic
phase are quite different from those of a conventional metal.
The main purpose of this article is to argue that this
is because these materials are strongly correlated
electron systems\cite{note} and that a 
model introduced by Kino and Fukuyama\cite{fuku}
may be the simplest possible theoretical model to describe them.
Hopefully the article will stimulate more experimental
and theoretical work on these exciting materials.
Readers  interested in a broader review of organic
superconductors should consult recent monographs
on the subject.\cite{ish,wos,williams}

\section{Experimental properties}

Organic molecular crystals with the
formula (BEDT-TTF)$_2$X consist of
conducting layers of BEDT-TTF molecules
sandwiched between insulating layers
of anions X.
The layered structure leads to highly anisotropic
electronic properties.
BEDT-TTF is a large planar molecule
and the different possible packing patterns
are denoted by different Greek letters \cite{science}.
The basic unit of the packing pattern
in the $\kappa$ phase is
a ``dimer'' consisting of two BEDT-TTF molecules
stacked on top of one another.
 Each dimer has one electron
less than a full electronic shell
because of charge transfer to the anions.


 The family $\kappa-$(BEDT-TTF)$_2$X
 has a particularly rich phase diagram as a function
of pressure, temperature, and anion 
X=Cu[N(CN)$_2$]Cl, Cu[N(CN)$_2$]Br,
and Cu(SCN)$_2$, as shown in Fig. 1.
A number of features of this diagram should be noted.
(i) Antiferromagnetic and
 superconducting phases occur next to one another.
(ii) At low pressures the metallic phase has properties
that are quite distinct from conventional metals.
(iii) The phase diagram is quite similar
to that of the cuprates if pressure is
replaced with doping.
The properties of the different
phases are now reviewed.

\subsection{Metal-insulator transition}

In  X=Cu[N(CN)$_2$]Cl
under moderate pressures (20-30 Mpa)
as the temperature is lowered there is
a first-order transition from an insulating
into a metallic phase.\cite{ito}
This transition also occurs at constant temperatures
between 10 and 30 K as the pressure is increased.
At ambient pressure
X= Cu(CN)$_3$  is a paramagnetic insulator.\cite{komatsu}
For pressures of 0.35 to 0.4 GPa there is 
a first order transition as the temperature is
lowered. (For 0.35  GPa the resistivity
decreases by five orders of magnitude at the 
transition).
The temperature at which the metal insulator
transition occurs increases with pressure.
It will later  be argued that this
metal-insulator transition is a Mott-Hubbard
transition driven by electron-electron interactions.
However, it needs to be checked whether there
is any structural change associated with this
transition. V$_2$O$_3$ undergoes a metal-insulator
transition. Since there is a jump in one of
the lattice constants at the transition
its origin could be structural rather 
than electronic.

\subsection{Insulating antiferromagnetic phase}

Evidence of antiferromagnetic
ordering below 26 K is provided 
in  X=Cu[N(CN)$_2$]Cl \cite{miyagawa}
 and deuterated X=Cu[N(CN)$_2$]Br \cite{kanoda3} by
the observed splitting of proton nmr  lines.
The magnetic moment estimated from
the magnitude of the splitting is $(0.4 -1.0)\mu_B$ per dimer.
This large moment suggests that there are strong
electron-electron interactions in these
materials.
Significant anisotropy is seen in the magnetic susceptibility
below 26 K.
Below 22 K there is weak ferromagnetic 
hysteresis with a saturation moment of
about $10^{-3} \mu_B$ per formula unit.\cite{welp}
This small ferromagnetic moment could be due to a slight canting of the
antiferromagnetic moments.\cite{manuel}
Antiferromagnetic resonance\cite{afmr}
 results are consistent with an easy plane
antiferromagnet with Dzyaloshinskii-Moriya interaction.
The antiferromagnetic phase becomes unstable
under moderate pressures of about 300 bar.
It can be stabilised and enhanced by
a magnetic field perpendicular
to the layers of the order of a few tesla.\cite{sushko2}

The observed phase diagram at the
boundary of the superconducting and insulating
phases is actually more complex than shown
in Fig. 1.
For example, coexistence of superconductivity and
antiferromagnetic phases has been observed.\cite{sushko,ito,kanoda3}
As the temperature is lowered a superconductor
to insulator transition is observed.



\subsection{Metallic phase}

Many of the properties of the metallic phase have
a temperature dependence that is quite
distinct from that of conventional metals.
Yet at low temperatures (less than about 30 K)
some properties are similar to those of a
conventional metal, but with a Fermi energy
of the order of 100 K.
This is almost an order of magnitude smaller
than predicted by band structure calculations.
Application of pressures of the order of 10 kbar
restores conventional metallic properties
over the full temperature range.
The details of different properties are now reviewed.

{\it Optical conductivity.}
Infrared\cite{eldridge1,eldridge2} and microwave\cite{dressel2} measurements 
of the frequency dependent conductivity $\sigma(\omega)$ of
X=Cu(SCN)$_2$, X=Cu[N(CN)$_2$]Br, and X=I$_3$\cite{tamura}
deviate from the Drude behavior found in conventional metals.
At room temperature $\sigma(\omega)$ is
dominated by  a broad peak around
300 or 400 meV (depending on the polarization
and anion X) with a width of about 150 meV.
Even down to 50 K no ``Drude-like'' peak at
zero frequency is present (see Fig. \ref{figcond}).
At 25 K the high energy peak decreases slightly in
temperature and a ``Drude-like'' peak but                  
can only be fit to a Drude form if
the scattering rate and effective mass
 are frequency dependent.\cite{eldridge1}


{\it Nuclear magnetic resonance.}
In a conventional metal, the Knight shift $K$ is proportional
to the density of states at the Fermi energy and  is independent
of temperature.\cite{slichter} In contrast, in
X=Cu[N(CN)$_2$]Br
 the Knight shift decreases significantly below about 50 K,
suggesting a suppression of the density of
states or ``pseudogap'' near the Fermi energy.\cite{desoto,mayaffre}
In conventional metals the NMR relaxation rate $1/T_1$
obeys Korringa's law:\cite{slichter}
$ 1/(T_1 T K^2)$ is 
independent of temperature and a universal number.
The relaxation rate is five to ten
times larger than expected from Korringa's law
 and is strongly temperature dependent:
$ 1/(T_1 T)$ has a peak in the range
10 to 50 K, depending on how
close the system is to the metal-insulator
transition.\cite{desoto,kawamoto,kawamoto2}
Above 50 K the relaxation rate is similar for
X=Cu[N(CN)$_2$]Cl,  Cu[N(CN)$_2$]Br
 and X=Cu(NCS)$_2$ \cite{kawamoto,kawamoto2}.
The above behavior has
some similarities to the temperature dependence of the nmr properties of
the underdoped cuprates.\cite{nmr}
As the pressure is increased to four  kilobars the NMR
properties become more like those of a conventional metal:
the Knight shift and 
$ 1/(T_1 T)$  become independent of temperature.\cite{mayaffre}

{\it Temperature dependence of the resistance.}
Unlike a conventional metal the resistance does
not monotonically increase with temperature.
It is a maximum at $T_{max}$ where
$T_{max} \simeq 100 $ K
for X=Cu(NCS)$_2$.
With increasing pressure $T_{max}$ increases
and the peak disappears at high pressures.\cite{kanoda3}
From about 30 K down to the superconducting
transition temperature the resistance 
decreases quadratically with temperature\cite{dressel}
\begin{equation}
 \rho(T) = \rho_0 + A T^2
\label{resist}
\end{equation}
Such a temperature dependence is characteristic
of metals    in which the dominant scattering mechanism 
is the interactions of the electrons with one
another and is observed in transition metals
and heavy fermions. 
In those systems
the Kadowaki-Woods rule\cite{kadowaki,woods}
relates the coefficient $A$ in (\ref{resist})
 to the linear coefficient for the specific heat, $\gamma$:
$A/\gamma^2=$constant. The constant is
$4.0 \times 10^{-13} \Omega$cm (mol/mJ)$^2$
for transition metals, and 
$1.0 \times 10^{-11} \Omega$cm (mol/mJ)$^2$
for heavy fermions and 
for transition metal oxides
near the Mott-Hubbard transition\cite{tokura}.
In the organics the ratio is 5 to 200 times larger than
predicted by this law\cite{haddon2,dressel}.
It is has been suggested that this discrepancy 
implies that the dominant scattering mechanism is
phonons.\cite{haddon2,weger}
In some strong coupling superconductors\cite{gurvitch}
and dirty metallic films\cite{sergeev}
a $T^2$ resistivity is observed.
If interference between electron-phonon
scattering and disorder is taken into account
theory predicts that $A \simeq \alpha \rho(0)/\theta_D^2$
where $\alpha \sim 0.01-0.1$, and $\rho(0)$
is the residual resistivity due to disorder.
Comparing to the data on 
X=Cu(SCN)$_2$
in Ref. \onlinecite{dressel}, the value of $A$
predicted by this expression is several orders 
of magnitude smaller than observed.


{\it Mean-free path and Mott's minimum conductivity.}\cite{weger2}
If the electronic mean-free path is less than a lattice 
constant than the idea of electronic transport
by electrons with well-defined wavevector is
no longer meaningful.\cite{mott}
Then the material should act as an insulator
  with a resistance  that drops with
increasing temperature;
to the contrary it remains a metal with an increasing resistance.
If the mean-free path is of the order of a lattice constant
then for a quasi-two dimensional Fermi surface
the intralayer conductivity will be approximately,
\begin{equation}
 \sigma_{min} \sim  {e^2 \over h} {B  \over c}
\label{min}
\end{equation}
where $c$ is the interlayer spacing and $B$ is a constant
of order one. For $c \sim 10 \AA$ this gives
$\sigma_{min} \sim 10^3 (\Omega {\rm cm})^{-1}$.
The intralayer conductivity of
 X=Cu(NCS)$_2$ 
 is comparable to
this at 30 K and is three orders of magnitude smaller at 100 K.\cite{dressel}
However, from 30 to 100 K, the resistance increases with temperature,
characteristic of a metal.
Hence, these materials can be classified as ``bad metals.''\cite{emery}

{\it Hall resistance.} 
In a conventional metal the Hall resistance
is a measure of the number of charge carriers
and is independent of temperature
above about $0.3 \theta_D$, where $\theta_D$
is the Debye temperature.\cite{hurd}
(Specific heat measurements\cite{andraka} suggest
$\theta_D \sim  $ 200 K in the $\kappa$ salts).
The Hall resistance of 
X=Cu(SCN)$_2$
is weakly temperature dependent above 80 K
but increases by a factor of three as
the temperature is decreased from 80 K
down to the superconducting transition temperature
(10 K).\cite{murata}
Hall measurements on 
X=Cu[N(CN)$_2$]Br and
X=Cu[N(CN)$_2$]Cl
at ambient pressure\cite{tanatar}
also found a strong temperature dependence
but were complicated by long-term sample resistance
relaxation.
Recent measurements on the
X=Cu[N(CN)$_2$]Cl
 salt under pressure also found a strong temperature dependence
\cite{sushko}.
 However, the ratio of the longitudinal
resistance $R_{xx}$ to the Hall resistance  $R_H  $ has
a simple quadratic temperature dependence up to 100 K:
\begin{equation}
 \cot \theta_H \equiv { R_{xx} \over R_H } = \alpha T^2 + C.
\end{equation}
$\theta_H$ is known as the Hall angle.
Similar behavior is found in the cuprate
superconductors\cite{ong2}.
It has been suggested that in the cuprates
that there are two distinct scattering times
and that these have a different temperature
dependence.\cite{and,col}
The Hall resistance involves the product of
the two scattering rates.
This hypothesis needs to be tested in the organics by checking if
$R_{xx}^2 / R_H$ is independent of temperature.

{\it Thermoelectric power.}
In a conventional metal the thermoelectric power is given by 
\begin{equation}
S = {\pi^2 \over 3 e} { k_B^2 T \over E_F }.
\end{equation}
where $k_B$ is Boltzmann's constant, $e$ is the electronic charge,
and $E_F$ is the Fermi energy.
The Fermi energy estimated from comparing
this expression to 
measurements\cite{yu}
on X=Cu[N(CN)$_2$]Br
is several times smaller than
predicted by H\"uckel band structure calculations.
Furthermore, the thermopower is not linear in
temperature above 30 K and has a peak around 100 K
(as does the resistivity) and is similar in magnitude
and temperature dependence to the cuprates\cite{zhou}.

{\it Specific heat.}
The low-temperature specific heat of 
X=Cu[N(CN)$_2$]Br
and X=Cu(SCN)$_2$ have been measured\cite{andraka2,andraka}
at ambient pressure and under a magnetic field
of 10 tesla (to destroy the superconductivity).
The specific heat coefficients $\gamma$ are
22 $\pm$ 3 mJ K$^{-2}$ mol$^{-1}$ and              
25 $\pm$ 3 mJ K$^{-2}$ mol$^{-1}$, respectively,
about 2.5 times larger than
predicted by H\"uckel band structure calculations.\cite{haddon}

{\it Magnetic susceptibility.}
In a conventional metal this is independent of
energy for temperatures much less than
than the Fermi energy.
For the $\kappa$ salts the susceptibility $\chi$ is almost independent
of temperature above 50 K but decreases slightly below
50 K.\cite{haddon}

{\it Wilson ratio.}
The dimensionless quantity $R \equiv 4 \pi^2 k_B^2 \chi(0)/(3 (g
\mu_B)^2 \gamma)$
where $g \simeq 2$ is the gyro-magnetic ratio
and $\mu_B$ is the Bohr magneton.
For a non-interacting Fermi gas $R=1$ and
for the Kondo model $R=2$.
For heavy fermion metals it is in the range one to three.\cite{wilkins}
For both X=Cu[N(CN)$_2$]Br
and X=Cu(SCN)$_2$ it is 1.5 $\pm$ 0.2.

{\it Magneto-oscillations.}
The temperature dependence of the amplitude of
Shubnikov - de Haas and de Haas - van Alphen 
oscillations can be used to deduce the
effective mass, $m^*$, of the charge
carriers associated with various orbits
on the Fermi surface.\cite{wos} 
Values obtained for the $\beta$ orbit
include $m^*/m_e = 5.4 \pm 0.1,
7.1 \pm 0.5,
 3.9 \pm 0.1$ for  
X=Cu[N(CN)$_2$]Br,\cite{mielke}
X=Cu(SCN)$_2$,\cite{harrison}                     
and X=I$_3$,\cite{schweitzer} respectively.
For the $\alpha$ orbit in X=Cu(SCN)$_2$
a value of 
$3.3 \pm 0.1$ has been obtained.  \cite{harrison,singleton,wosnitza2,oshima2}
Under a pressure of 20 kbar this effective mass
decreases to 1.5$m_e$.\cite{singleton}
The rapid decrease in the effective mass
with increasing pressure
was correlated with the decrease
in the superconducting transition temperature.
A cyclotron resonance measurement
(which should measure the bare mass
and not the effective mass\cite{kohn}) on
 X=Cu(SCN)$_2$
gives a mass of 1.2$m_e$.\cite{hill}


The measured effective masses are 
two to four times larger than predicted by 
H\"uckel band-structure calculations that ignore the interactions 
between the electrons.
Calculations using the local-density approximation
(LDA) for X=Cu[N(CN)$_2$]Br
give a band width a factor
of two smaller than H\"uckel.\cite{Ching}


The fact that the effective mass, specific heat,
and magnetic susceptibility are significantly
larger than predicted by band structure calculations
suggest that the band width is narrowed by
many-body effects.

\subsection{Superconducting phase}

There is increasing evidence that, like in
the cuprates,\cite{scalapino} 
the pairing of electrons in the superconducting     
state involves a different symmetry state 
than in conventional metals.
Below I briefly review the evidence, pointing
out where controversy exists.

{\it Nuclear magnetic resonance.}
The NMR relaxation rate $1/T_1$ does
not have a Hebel-Slichter peak below $T_c$.
At low temperatures, $ 1/T_1 \sim T^3$,\cite{desoto,kanoda}
instead of the exponentially activated 
behavior found in conventional superconductors.
This power law behavior suggests there are nodes
in the gap.

{\it Specific heat.}
At low temperatures the electronic specific heat of 
 X=Cu[N(CN)$_2$]Br 
 has a power law 
dependence on temperature, in contrast to
the exponentially activated dependence of
conventional superconductors.
In a magnetic field $B$ perpendicular to the
layers the electronic specific heat 
is linear in temperature with
coefficient $\gamma \sim B^{1/2}$.\cite{nakazawa}
This is the behavior predicted
for a clean two-dimensional superconductor
with nodes in the gap.\cite{volovik} 
In contrast, in a conventional superconductor $\gamma \sim B$.

{\it  Magnetic penetration depth.}
There is controversy about its 
temperature dependence.
dc magnetisation measurements on X=Cu(SCN)$_2$
have been fitted to the exponentially
activated temperature dependence
expected for an s-wave superconductor.\cite{lang}
In contrast, muon spin rotation ($\mu$SR) measurements
on X=Cu(SCN)$_2$ and  X=Cu[N(CN)$_2$]Br 
 show a linear temperature
dependence with temperature.\cite{le}
Other $\mu$SR measurements are consistent
with an s-wave order parameter.\cite{harshman}
Surface impedance measurements on 
on X=Cu(SCN)$_2$ and  X=Cu[N(CN)$_2$]Br 
give a penetration depth whose temperature dependence
is consistent with s-wave pairing.\cite{dressel2}
Furthermore, the real part of the conductivity
versus temperature has a `coherence' peak 
just below $T_c$, as in conventional superconductors.

{\it Upper critical field $H_{c2}$.}
The $H_{c2}$ versus temperature curve
for X=Cu(SCN)$_2$ 
shows upper curvature near $T_c$.
The large slope near $T_c$
suggests that  
at low temperatures  $H_{c2}$ could exceed the Pauli limit
(the field at which BCS theory predicts
singlet Cooper pairs should be broken).\cite{kwok,oshima}

{\it Vortex dynamics.}
Vibrating reed studies of 
 X=Cu[N(CN)$_2$]Br 
in magnetic fields less 
than one tesla suggest anomalous vortex dynamics
and the possibility of unconventional superconductivity.\cite{delong}

\section{Theory}

I now argue that a theoretical model introduced by Kino
and Fukuyama\cite{fuku} 
captures enough of the essential physics
of these materials to describe them, at least at the semi-quantitative
level.
Insight into the model is obtained by considering its
various limits and what is known about similar models.

\subsection{Band structure}

The arrangement of the BEDT-TTF molecules
in the $\kappa$ crystal structure and the dominant
intermolecular hopping integrals are shown in Figure
 \ref{figx}.
  Table I shows values for these integrals
 found from various  quantum chemistry calculations for different anions.
 The intradimer hopping $t_{b_1}$
is more than two times larger than the interdimer integrals,
 $t_p$  and $t_q$  along the $c+a$
direction, and $t_{b_2}$ along the $c$ direction.
The dimer bonding and antibonding orbitals are split by 
approximately $2t_{b_1}$ and so
the mixing between bonding and antibonding orbitals
can be neglected and the interdimer hoppings
can be treated as a perturbation.
Focusing on antibonding orbitals, each dimeric site has two nearest
neighbors
along the $c$ direction (interaction $t_1 \equiv t_{b_2}/2$), and 
four next-nearest neighbors
along the $c \pm a$ directions (interaction $t_2 \equiv (t_p+t_q)/2$).
The dimer model is shown in Fig. \ref{fig2}
(compare Figures 1 and 6 in Ref. \onlinecite{tamura} and Figure
  13 in Ref. \onlinecite{fuku} )
Note that each lattice site has hopping to six neighbours
and so the lattice has the same co-ordination as the triangular
lattice.
Since each dimer has three electrons (or one hole),
 the upper (conduction) band is half-filled.

If the interactions between electrons are neglected
(which it will be argued below is not justified) then
the model is just a tight-binding model and
the dispersion relation for the half-filled band is
\begin{equation}
 E(k_a, k_c) = 2 t_1   \cos \left(k_c c\right) + 4 t_2         
 \cos {\left(\frac{k_a a}{2}\right)}
                \cos {\left(\frac{k_c c}{2}\right) }
\label{tight}
\end{equation}
This band structure reproduces the main features
of bands calculated 
using the extended H\"uckel approximation\cite{girlando,gei91}.
Bands calculated using the local-density approximation
(LDA) for  X=Cu(NCS)$_2$ and 
X=Cu[N(CN)$_2$]Br
are also similar apart from
an overall bandwidth narrowing.\cite{Ching}
The band structure (\ref{tight}) has been used to model
the results of Shubnikov - de Haas oscillations
measurements\cite{singleton} on 
X=Cu(NCS)$_2$ and reflectance measurements\cite{tamura}
on X=I$_3$.

\subsection{Estimates of the Coulomb repulsion}
\label{estimate}

To test whether these materials can be 
described by a tight-binding model
which neglects electron-electron interactions
the effective Coulomb repulsion
between two electrons in the anti-bonding orbital
on a BEDT-TTF dimer is now estimated.
Quantum chemistry calculations\cite{Fortu,okuno,castet}
 estimate that the Coulomb repulsion 
between two electrons on a {\it single} BEDT-TTF molecule $U_0$ 
is about 4-5 eV.
These calculations neglect the screening effects present in 
the solid state and so the actual $U_0$ may actually be smaller.
$U_0$ has been estimated 
from optical experiments to be 0.7 eV
in (BEDT-TTF)HgBr$_3$\cite{tajima}
and in $\alpha^\prime$-(BEDT-TTF)$_2$Ag(CN)$_2$
to be  1.3 eV.\cite{friend}

If $E_0(N)$ denotes the ground state energy of $N$
electrons on a dimer then\cite{ash,fuku} $E_0(2) = { 1 \over 2} ( U_0 - ( U_0^2
 + 16 t_{b_1}^2)^{1/2} ),$
$E_0(3) = U_0 -t,$  and $E_0(4) = 2 U_0.$
The effective Coulomb repulsion $U$ between two electrons 
on a dimer is then 
\begin{equation}
U = E_0(4) + E_0(2) - 2 E_0(3) =
2t_{b_1} + { 1 \over 2} ( U_0 - ( U_0^2 + 16 t_{b_1}^2)^{1/2} )
\label{dimerU}
\end{equation}
If $U_0 >> 4 t_{b_1}$
the effective repulsion between two holes 
on a   dimer is approximately $2 t_{b_1}$ (See Figure \ref{figU}).
However, if $ U_0 \sim 4 t_{b_1} \sim $ 1 eV
then $U \sim $ 0.3 eV.
The optical conductivity data shown in Fig. \ref{figcond}
is consistent with this estimate.
This value of $U$ is comparable to the band width of the tight binding
model and so suggests that correlation effects are important.

\subsection{A minimal model: the dimer Hubbard model}

Based on the above discussion we suggest
that a minimal theoretical model to describe
the $\kappa$-(BEDT-TTF)$_2$X crystals is
the following Hamiltonian,\cite{fuku}
\begin{equation}
 H=
-t_1\sum_{<{\bf{ij}}>,\sigma}(c^{\dagger}_{{\bf{i}},\sigma}
c_{{\bf{j}},\sigma}+h.c.)
-t_2\sum_{<{\bf{in}}>,\sigma}(c^{\dagger}_{{\bf{i}},\sigma}
c_{{\bf{n}},\sigma}+h.c.)
+U \sum_{{\bf{i}}}(n_{{\bf{i}} \uparrow}-{1 \over 2})( n_{{\bf{i}}
\downarrow}- { 1 \over 2})
+\mu\sum_{{\bf{i}},\sigma}n_{{\bf{i}}\sigma} ,
\label{hubb}
\end{equation}
\noindent where ${\rm c^{\dagger}_{{\bf{i}},\sigma} }$ creates a hole at
dimer site ${\rm {\bf i } }$
with spin projection $\sigma$, ${\rm n_{{\bf{i}}\sigma} }$ is the 
hole number operator.
The sum ${\rm \langle {\bf{ij}} \rangle }$
runs over pairs of  nearest neighbor lattice sites
in the horizontal direction, and the sum
${\rm \langle {\bf{in}} \rangle }$
runs over pairs of lattice sites along the diagonals.
$U $ is the on-site Coulombic repulsion on a single dimer
(estimated in the previous section), 
and $\mu$ is the chemical potential.
At half-filling  $\mu$ may be non-zero due
to the absence of particle-hole symmetry if both 
$t_1$ and $t_2$ are non-zero.

If there are strong correlations on each dimer then
the amplitude for hopping between neighouring dimers,
$t_1$ and $t_2$, are not simply given by
$t_1 = t_{b_2}/2$ 
and  $t_2 = (t_p+t_q)/2$.
The correct amplitude involves the overlap
of a one hole state with the anti-bonding
orbital creation operator acting on a
two-hole state.\cite{correct}
A straight-forward calculation gives
\begin{equation}
{t_1  \over  t_{b_2}}
= {t_2 \over  (t_p+t_q)}   
= { 1 \over 2 \sqrt{2}} \left( \cos \theta - \sin \theta \right)
\label{suppr}
\end{equation}
 where 
\begin{equation}
\tan \theta = 
  {U_0 \over 4t_{b_1}} -
 \left( \left({U_0 \over  4 t_{b_1}}\right)^2 + 1 \right)^{1/2} \end{equation}
If $U_0 >> 4 t_{b_1}$
then correlations on the dimer reduce the
inter-dimer hopping by a factor of $1/\sqrt{2}$.

Various limits of the model (\ref{hubb}) are now considered
leading to the speculative phase diagram shown in Fig. \ref{fig3}.

An important property of this model is that at
half-filling a non-zero $U$ 
is required for an insulating phase.
A finite $U$ is required because if $t_1$ and $t_2$ are both non-zero
the nesting of the Fermi surface is not perfect.
In the Hartree-Fock approximation
Kino and Fukuyama\cite{fuku} found
that for $2t_1 = t_2$, the critical value is 
 $U_c/t_2 \simeq 3.$

If $t_1=t_2=t$ the model becomes the Hubbard model on a triangular lattice,
after a rescaling of the lattice constants.
Although there have been a number of studies of this model
at half filling, no clear consensus of the phase diagram
has been reached except that the ground
state should be a paramagnetic metal for small
$U/t$ and an insulator for large $U/t$.
The model has been studied by techniques including
the Hartree-Fock
approximation (which predicts $U_c/t \simeq 5)$,\cite{sarker} the slave boson
technique,\cite{gazza}
and exact diagonalization of small clusters.\cite{pecher}
The slave boson technique
predicts that between the paramagnetic metal
and insulating antiferromagnetic phase that there
is semimetallic linear commensurate spin-density-wave
phase for $6.9 < U/t < 7.8.$\cite{duffy}
In order to allow for the slightly 
different band structure associated with the
lattice considered here these critical values
of $U$ must be divided by $\sqrt{3}$.
Both the Hartree-Fock and slave-boson
(mean-field) approximations tend to underestimate
the effect of fluctuations and so the actual
critical $U$ may be larger than the above values.

The estimate of $U$ in the previous
section and the parameter values given in Table I
for various
$\kappa-$(BEDT-TTF)$_2$X
crystals suggest that $U/(t_1 + t_2) \sim 3-6$ and
so these materials may be 
close to the metal-insulator transition.

\subsection{Strong coupling limit for the insulating phase}

If $U \gg t_1,t_2$ then the ground state is an insulator
and a standard strong-coupling expansion\cite{auerbach}
for the Hamiltonian (\ref{hubb})
implies that the spin degrees of freedom are
described by the spin-${1 \over 2}$ Heisenberg model
\begin{equation}
\label{ham}
    H = J_1 \sum_{\rm \langle {\bf ij} \rangle} \bbox{S}_{\bf i} \cdot
 \bbox{S}_{\bf j}
  + J_2 \sum_{\rm \langle {\bf in} \rangle} \bbox{S}_{\bf i}
 \cdot \bbox{S}_{\bf n} 
\end{equation}
where $\bbox{S}_{\bf i}$ denotes a spin operator
on site ${\bf i}$, $J_1 = t_1^2/U$ and $J_2 = t_2^2/U.$ 
Again, the sum ${\rm \langle {\bf{ij}} \rangle }$
runs over pairs of  nearest neighbor lattice sites
in the horizontal direction, and the sum
${\rm \langle {\bf{in}} \rangle }$
runs over pairs of lattice sites along the diagonals.
$J_1$ and $J_2$ are competing interactions
leading to magnetic frustration.
The parameter values in Table I suggest that
$J_1/J_2 \sim 0.3 - 1$ 
and so magnetic frustration will
play an important role in these materials.
This model has some similarites             
to a model considered for the layered
titanate Na$_2$Ti$_2$Sb$_2$O 
and with possible spin liquid ground states.\cite{singh}

Insight can be gained by considering various limits of this model.
If $J_1=0$ then the model reduces to
the Heisenberg model on a square lattice.
At zero temperature
there will be long range Neel order with
a magnetic moment of $0.6 \mu_B$.\cite{auerbach}
If $J_1$ is non-zero but small it will introduce
a small amount of magnetic frustration
which will reduce the magnitude of the
magnetic moment in the Neel state.
The spin structure is shown in Fig. \ref{fig4}
the same as that
proposed for the antiferromagnetic phase
of deuterated             
X=Cu[N(CN)$_2$]Br
based on nmr\cite{kanoda3}.
As the ratio $J_1/J_2$ varies the wave vector
associated with the antiferromagnetic order
may also vary.
There is the possibility of commensurate-incommensurate 
transitions.

If $J_1=J_2$ then the model reduces to
the Heisenberg model on a triangular lattice.
There has been some controversy about the
ground state of this model. Anderson\cite{anderson2}
originally suggested
that the ground state was a ``spin liquid''
with no long-range magnetic order.
However, recent numerical work suggests that
there is long-range order but the quantum
fluctuations are so large
due to the magnetic frustration that the magnetic moment
may be an order of magnitude smaller than
the classical value.\cite{singh2}

If $J_2=0$ then the model reduces to
a set of decoupled Heisenberg antiferromagnetic chains
which do not have long-range order.\cite{auerbach}
If $J_2$ is non-zero but small 
the chains are weakly coupled. 
The case of only two
chains corresponds to the
``zig-zag'' spin chain which is
equivalent to a single chain with
nearest-neigbour and next-nearest neighbour exchange,
$J_2$ and $J_1$, respectively.
(Unfortunately, the definition of
$J_1$ and $J_2$ is the opposite of that
normally used).
This spin chain has been extensively studied and
is well understood.\cite{bursill}
In the limit of interest here,
$J_2  \ll J_1$, there is a gap in the
spectrum $\Delta \sim \exp( -{\rm const.} J_1/J_2)$
and there is long-range dimer order.
We speculate that this ``spin gap''
is still present in the many chain limit.

Based on the above discussion a schematic phase diagram
for the model (\ref{hubb}) at half filling and 
at zero temperature has been drawn in
Fig. \ref{fig3}. The possibility of
superconductivity near the metal-insulator
transition is suggested only on the basis
of comparison with the experimental phase diagram.

\subsection{The effect of pressure}

\begin{figure}
\centerline{\epsfxsize=9cm  \epsfbox{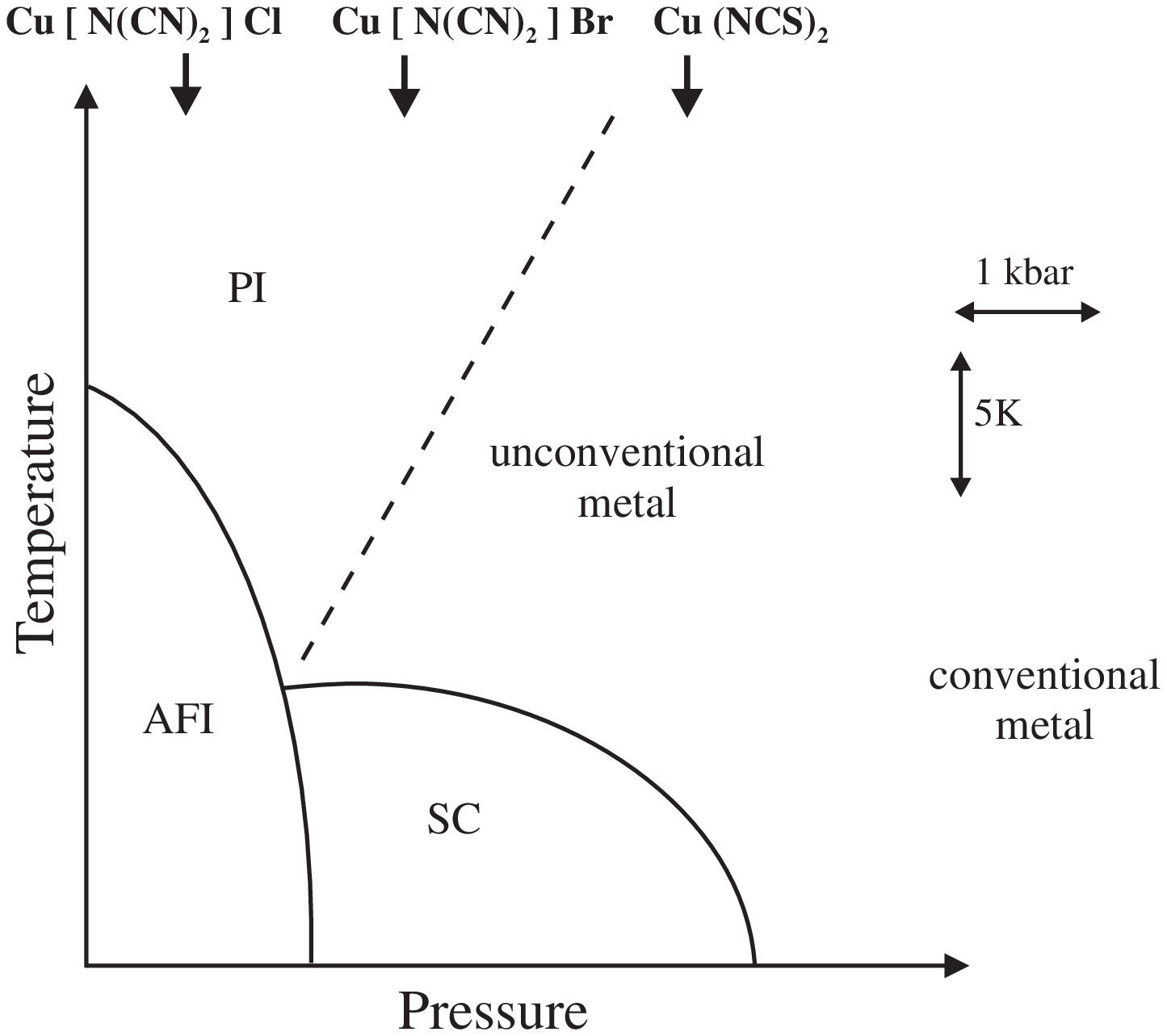}}
\vskip 0.3 cm
 \caption{
Schematic phase diagram of the 
$\kappa-$(BEDT-TTF)$_2$X
family of organic conductors.\protect\cite{fuku,kanoda3,wzietek}
AFI denotes an insulating antiferromagnetic phase,
PMI an paramagnetic insulating phase,
and SC a superconducting phase.
The phase transition between the paramagnetic
phases is first order and denoted by
the dashed line which ends at a
second order critical point.
The arrows denote the location
of at ambient pressure
of materials with different anions X.
As the pressure increases the unconventional
properties of the metallic phase disappear.
The above phase diagram is qualitatively
similar to that of the cuprate superconductors
with doping playing the role of pressure.
\label{figphase}} \end{figure}

A very important challenge is to understand
how varying the pressure changes the ground 
state of the system. The experimental
phase diagram shown in Fig. \ref{figphase}
certainly suggests that pressure reduces the
electronic correlations.
 In most strongly
correlated systems pressure increases the 
band width by more than the Coulomb repulsion
and so reduces the effect of correlations.
The effect of pressure in this system is
more subtle because the Coulomb repulsion $U$
is related to the hopping integral $t_{b_1}$
(see equation (\ref{dimerU}))
which will increase with pressure
and the ratio $t_1/t_2$ may vary with pressure
and could drive the metal-insulator transition.
Rahal et al.\cite{rahal} measured the crystal
structure of
 X=Cu(NCS)$_2$ 
at room temperature and pressures of 1 bar
and 7.5 kbar.
They then performed  H\"uckel calculations
of the hopping integrals for the corresponding 
crystal structures. The results, given in Table I,
suggest that the amount of magnetic frustration
decreases as the pressure increases.
In contrast to the work of Rahal et al. 
a more heuristic approach was taken in an earlier study 
by Campos et al.\cite{campos}
In order to fit experimental data for the pressure
dependence of the Fermi surface area deduced from
magneto-oscillation frequencies, they made  assumptions
about how the BEDT-TTF molecules move relative
to one another as the pressure varies.
They found an increase in $t_1/t_2$ with pressure.


Assuming that the Coulomb repulsion on a single BEDT-TTF 
molecule $U_0$ does not vary significantly with pressure
then equation (\ref{dimerU}) implies
\begin{equation}
{ d U \over d P} = 
2 { d t_{b_1} \over d P} \left( 1 -  { 4 t_{b_1} \over 
( U_0^2 + 16 t_{b_1}^2)^{1/2} }\right)
\label{pressure}
\end{equation}
If $U_0 >> 4 t_{b_1}$
then $U$ will increase with pressure at about
the same rate as $t_{b_1}$ which is similar to
the rate at which $t_1$ and $t_2$ increase.
This could mean that the correlations change little 
with pressure.
However, if $ U_0 \sim 4 t_{b_1}$ then the
second term in (\ref{pressure}) reduces the 
rate at which $U$ increases and so 
correlation effects can decrease with pressure.
The uncertainty in the variation of the parameters
with pressure is sufficiently great that more
work is required before definitive answers     
can be obtained about this important question
of how the ground state changes with pressure.

\subsection{Dynamical mean-field theory for the Hubbard model}
\label{dmft}

Some insight into the physics of the organics,
particularly the existence of a low-energy scale
in the metallic phase, can be gained by
briefly reviewing
progress that has been made in the past few years in understanding
the Mott-Hubbard metal-insulator transition.\cite{review}
This has been done by
studying the Hubbard model in a dynamical mean-field approximation which
becomes exact in the limit of either
large {\it lattice connectivity}
or {\it spatial dimensionality.}
This approximation maps the problem onto a single impurity
problem that must be solved self-consistently.
While time-dependent fluctuations are captured
by this approximation spatially-dependent fluctuations
are not captured.
Some important physics that emerges from these studies is that there is 
an important low-energy scale $T_0$ which is much smaller than
the half-bandwidth $D$ and the Coulomb repulsion $U$.
This energy scale is the analogue of the Kondo
temperature for the impurity problem.
If there is significant magnetic frustration
the Neel temperature for antiferromagnetic ordering
is suppressed. Then at half filling and
at fixed temperature $ T < T^* \simeq 0.05 D$
as $U$ is increased 
there is a first-order phase transition from
a paramagnetic metallic phase to a paramagnetic insulating phase
at $U = U_c \sim 3 D$.

In the metallic phase the density of states $\rho(\omega)$
contains peaks at energies $\omega = -U/2$ and $+U/2$
 which correspond to the lower and upper
Hubbard bands, respectively and involve incoherent states.  These peaks
are broad and have width of order $D$.  At temperatures less than
$T_0$ a new quasi-particle
peak with width of order $T_0$ forms at $\omega=0$  (Figure 39 in Reference 
\onlinecite{review}). 
The spectral weight of the peak vanishes as the
metal-insulator transition is approached.
The quasi-particle band  involves coherent states that form a Fermi liquid.
This means that the self energy 
$\Sigma (\omega)$
is independent of momentum
has the energy dependence,
${\rm Re} \Sigma (\omega)
 = U/2 + (1 - 1/Z) \omega + O (\omega^3)$
 and ${\rm Im} \Sigma (\omega) =  - B    \omega^2 +
O (\omega^4)$. $Z$ is the quasiparticle residue
and is related to the effective mass $m^*$ by
$m^*/m_e = 1/Z$.
 As the metal-insulator transition is approached $T_0 \to 0$
and $ Z \to 0$.\cite{fisher}
The consequences of the above picture for the temperature
dependence of different physical quantities for metals in close proximity
to the metal-insulator transition are now
briefly described.

{\it Optical conductivity.}\cite{rkk,cox}  For temperatures $T$
larger than $T_0$ the optical conductivity is small near zero-frequency and has
a broad peak at $\omega=U$,
corresponding to transitions between the lower
and upper Hubbard bands.  For $T < T_0$ a
 Drude-like peak forms at zero-frequency
and there is a new-shoulder at $\omega \sim U/2.$ 
 The latter peak corresponds to
transitions from the quasi-particle band to the upper Hubbard band and
from the lower Hubbard band to the quasi-particle band.
  The total spectral weight increases as the temperature decreases.
 This picture has
been used to explain the temperature dependence of the optical conductivity
of vanadium oxides.\cite{rkk,rim,rkk2}
If this picture is used to explain the data 
shown in Fig. \ref{figcond}
then the Coulomb repulsion $U$ is estimated
to be about 250 to 350 meV.
This is roughly consistent with the estimates
in Section \ref{estimate}.

{\it Resistivity.} 
The temperature dependence of the
 dc resistivity $\rho(T)$ is not monotonic.
It  has a peak at a temperature of the order of
$T_0$, and decreases rapidly below the peak. (See Figure 9
in Reference \onlinecite{cox}, Figure 2 in
Reference \onlinecite{majumdar2}, and the inset of
Figure 2 in Reference \onlinecite{rkk}).

{\it Mean-free path.}
Since coherent quasiparticles only exist for $T < T_0$ it is not required that
the mean-free path be larger than a lattice constant at high
temperatures.  Thus, it is possible
for the resistivity to increase with temperature and be larger than the
Mott minimum resistivity. 

{\it Fermi liquid regime.}
The properties of the low temperature range ($T < T_0$) in
which the metal has Fermi liquid properties is now considered.

{\it Specific heat.}  In the Fermi liquid regime the
 linear coefficient is\cite{fisher}
\begin{equation}
\gamma = { 2 \pi k_B^2 \over 3 Z D}
\label{specific}
\end{equation}
where $Z=0.9 (1 - U/U_c)$ is the quasi-particle weight.
The effective Fermi energy is $ \epsilon_F^* = Z D$.

{\it Effective mass.}
This is related to the quasiparticle weight $Z$ by
\begin{equation}
{m^* \over m_e} = { 1 \over Z}
\label{mstar}
\end{equation}
and diverges at the transition,
consistent with the Brinkman-Rice scenario\cite{br}
for the metal-insulator transition.

{\it Resistivity.} 
At low temperatures $\rho(T)
\sim AT^2$, characteristic of a Fermi liquid.
Near the Mott-Hubbard transition the coefficient $A$ is
related to the specific heat coefficient $\gamma$ by
$A/\gamma^2 = (2.3 a) \times 10^{-12} $ ohm cm (mol/mJ)$^2$
where $a$ is the lattice constant in $\AA$
of a three-dimensional system.\cite{fisher}

{\it Magnetic susceptibilities.}
The uniform susceptibility (in units in which $g \mu_B = 2$) is
\begin{equation}
\chi(q=0) = { 1 \over 0.45 \epsilon_F^* + J}
\label{susc0}
\end{equation}
where $J \equiv D^2 / 2 U$ is the magnetic exchange energy.
Note that this does {\it not} diverge at the transition.
In contrast, the local susceptibility
\begin{equation}
\chi_{loc} \equiv \sum_q \chi(q) = {4.7  \over \epsilon_F^*}
\label{susc}
\end{equation}
The local susceptibility can be related to the
spin-echo decay rate $1/T_{2G}$.\cite{decay}

{\it The Wilson ratio.}
  This is only universal for the local susceptibility.
\begin{equation}
R \equiv 
{ \chi_{loc} / \chi_{loc}^{free}
\over  \gamma     / \gamma^{free} } =   2.8
\label{wilson} 
\end{equation}
where $\chi_{loc}^{free}$ and $\gamma^{free}$ are the values
in the absence of interactions.

{\it The nmr relaxation rate.}
The spin relaxation rate $1/T_1$ 
depends linearly on temperature
\begin{equation}
{ 1 \over T_1 T } = \lim_{ \omega \to 0} { \chi''_{loc}(\omega)
\over \omega} = {12.5 \over (\epsilon_F^*)^2}
 \label{korringa} \end{equation}
This could be combined with (\ref{susc})
to give a generalised Korringa-type relation.
However, this is not directly relevant to nmr
since the Knight shift $K$ is proportional to the
uniform susceptibility, not the local susceptibility.
The Korringa ratio $1/(T_1 T K^2)$ 
will diverge at the transition.

I am unaware of calculations of the complete
temperature dependence of $1/(T_1 T )$,
the thermopower, and Hall resistance.
However, these quantities have been calculated
for a doped Mott insulator.
It was found\cite{jarrell} that $1/T_1$ versus
 temperature had a maximum at a temperature of the order of $T_0$.
 The thermopower versus temperature curve\cite{jarrell} has a
maximum at a temperature of the order of
$T_0$ at which its magnitude is of order 40 $\mu$V/K.
The  Hall resistance
is temperature dependent.\cite{jarrell,majumdar}  The Hall angle has an
approximately quadratic temperature dependence.\cite{jarrell}

Luttinger's theorem is obeyed:  the area of the
Fermi surface is the same as in the absence of interactions.\cite{muller}
This means that even though there are strong
electron-electron interactions, H\"uckel band structure
calculations (which ignore the interactions)
may still be able to predict the Fermi surface area
correctly. It has often been found for the organics
that H\"uckel       calculations can produce Fermi surface areas
in agreement with the frequency of magneto-oscillations.\cite{wos}
This success should {\it not} be used to argue
that the electron-electron interactions are weak.

In conclusion, many of the above properties 
are similar to those of the metallic phase
of the $\kappa$-(BEDT-TTF)$_2$X salts,
particularly the presence of a low-energy scale 
below which Fermi liquid behavior is observed.
The dynamical mean-field theory results for
temperature and frequency dependences for
the conductivity appear to be 
very similar to experimental results.
Two notable differences exist between
the dynamical mean-field theory results and
properties of the $\kappa$-(BEDT-TTF)$_2$X salts.
First, the calculated  value of the
ratio $A/\gamma^2$ is 5 to 200 times smaller than observed.
Second, the nmr relaxation rate at low
temperatures is not linear in temperature,
as predicted by (\ref{korringa}).
It remains to be seen if these discrepancies 
occur because for these quantities the
two-dimensionality of the organics 
becomes important.

\section{Future directions}

In conclusion, the experimental properties
of this class of organic superconductors
have been reviewed with an emphasis on
similarities to the cuprates and
how the metallic phase has properties quite distinct 
from a conventional metal.
Quantum chemistry calculations can be used to
justify a minimal microscopic model:
a Hubbard model on an anistropic triangular lattice.
Appropriate parameters for this model suggest
that their is substantial magnetic frustration
and the system is close to a metal-insulator transition.
Consequently, recent results for the Mott-Hubbard transition
suggest that this model can explain the unconventional
metallic properties.
However, there are many details to be worked
out and so below some specific suggestions
are made for future theoretical and experimental work.

{\it Theory.}

Based on the above discussion Figure \ref{fig3}
shows a speculative phase diagram for the Hubbard model
(\ref{hubb}) at half-filling and zero temperature.
The challenge is to determine the actual phase diagram
using the arsenal of theoretical techniques
that have been developed over the past decade to
attack Hubbard models for the cuprates.
Numerical techniqes that could be used include
exact diagonalization\cite{dagotto} and
quantum Monte Carlo.\cite{dagotto,gubernatis}
Analytic techniqes that could be used include
the fluctuation-exchange approximation\cite{bickers}
and slave bosons.\cite{wen}
Insights might be gained from using ideas
from gauge theory of fluctuations\cite{gauge}
and the SO(5) unified theory of superconductivity
and antiferromagnetism.\cite{zhang}
The most important question is whether this Hubbard model
can produce superconductivity.
This is controversial for the square lattice model
away from half-filling.\cite{gubernatis}
Other questions of interest include:
Does the metallic phase have non-Fermi liquid properties?
Are spin liquid phases possible?

It was shown in Section \ref{dmft}
that dynamical mean-field theory can provide
important insights into the organics.
However, the connection with the model Hamiltonian
(\ref{hubb}) needs to be made more precise by working
out the appropriate lattice of large connectivity
or dimensionality to study.
Possibilities include the infinite connectivity limit
of the Bethe lattice
with nearest neighbour and next-nearest neighbour hopping
and the two-sublattice frustrated model.\cite{review}
In both these models the degree of magnetic frustration
can be varied.  Then        
the complete temperature dependence of resistivity,
Hall resistance,
thermopower, magnetic susceptibility, and nmr relaxation
rate should be calculated and compared to the organics.

Finally, as discussed in Section \ref{pressure}
a better understanding is needed of how
the model parameters vary with pressure.

{\it Experiment.}

The organics have a number of
advantages over the cuprates for experimental
study. First, since pressure and not doping
is used to  tune through the metal-insulator transition
(compare Figure \ref{figcup})
only a single sample is needed to study the
transition. Second, the lower superconducting transition
temperature and high purity samples mean that
 magneto-oscillations provide a powerful probe of
the Fermi surface.
On the other hand, the single crystals are sufficiently small
that neutron scattering is difficult and the first studies
on the organics have only recently been performed.\cite{pint}
Further studies could probe the magnetic fluctuations.
X=Cu[N(CN)$_2$]Br
lies close to the metal-insulator
transtion\cite{kawamoto2}
and so is the ideal candidate for further study.
More systematic experimental studies of the metallic state
and its variation with pressure are needed. 
 Extensive transport and thermodynamic
measurements should be made on the {\it same} single crystal.
Of particular interest is how
much of the low-temperature data can be explained
within  a Brinkman-Rice picture\cite{br} of  a Fermi liquid 
with an effective mass that diverges at the metal-insulator
transition.

By applying uniaxial stress along the $\bf c$-axis 
(in the conducting plane)
it should be possible to vary the ratio of
$t_1$ to $t_2$ and thus vary the
magnetic frustration.
A few uniaxial stress experiments\cite{campos2,kus,kund}
 have been performed.
One did find that for stress in the plane 
the resistivity peak and superconducting
transition temperature were enhanced.\cite{kus}
More extensive measurements are needed.

X=Cu$_2$(CN)$_3$ is particularly interesting because
in the insulating phase at ambient pressure there is no evidence of
antiferromagnet ordering at low temperatures.\cite{komatsu}
This should be checked with nuclear magnetic resonance.
The magnetic susceptibility decreases rapidly below 10 K.
Is this due to a spin liquid ground state?
The parameter values in Table I suggest that this material
has the largest amount of magnetic frustration of the listed
materials.

\begin{figure}
\centerline{\epsfxsize=9cm  \epsfbox{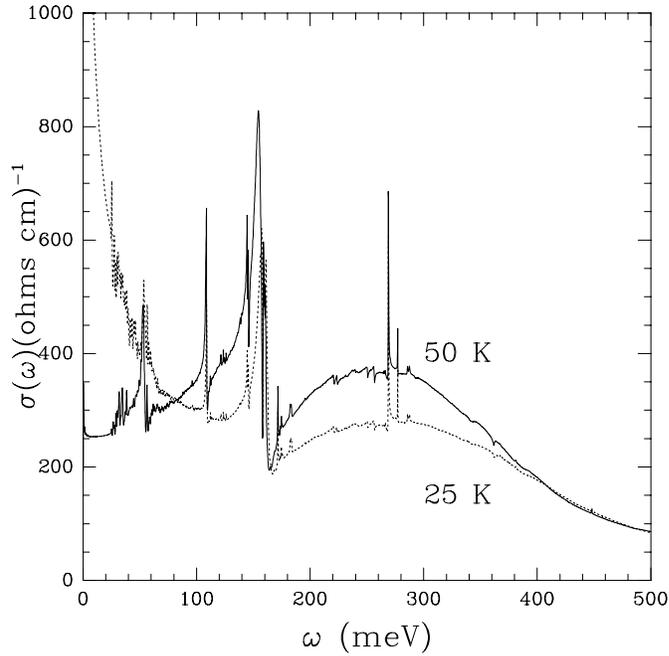}}
\vskip 0.3 cm
\caption{
Non-Drude behavior of the frequency-dependent
conductivity $\sigma(\omega)$ of 
X=Cu[N(CN)$_2$]Br.
The data is from Reference \protect\onlinecite{eldridge1}
and shows the strong temperature dependence of the
low-frequency conductivity.
At 25 K and below $\sigma(\omega)$ has a Drude-type
peak at zero frequency (see the dashed curve). In contrast, this peak is
completely absent at 50 K and above (see the solid curve).
The broad peak around 300 meV
can be identified with transitions between the
lower and upper Hubbard bands. The very sharp
features are due to infra-red active phonons.
The data shown is for the electric field parallel to
the {\bf a}-axis. Similar data  is
obtained for the electric field parallel to
the {\bf c}-axis except the broad peak is
around 400 meV.\protect\cite{eldridge1}
(Both the {\bf a}-axis and the {\bf c}-axis
are in the plane of the layers).
\label{figcond}} \end{figure}
\begin{figure}
\centerline{\epsfxsize=9cm  \epsfbox{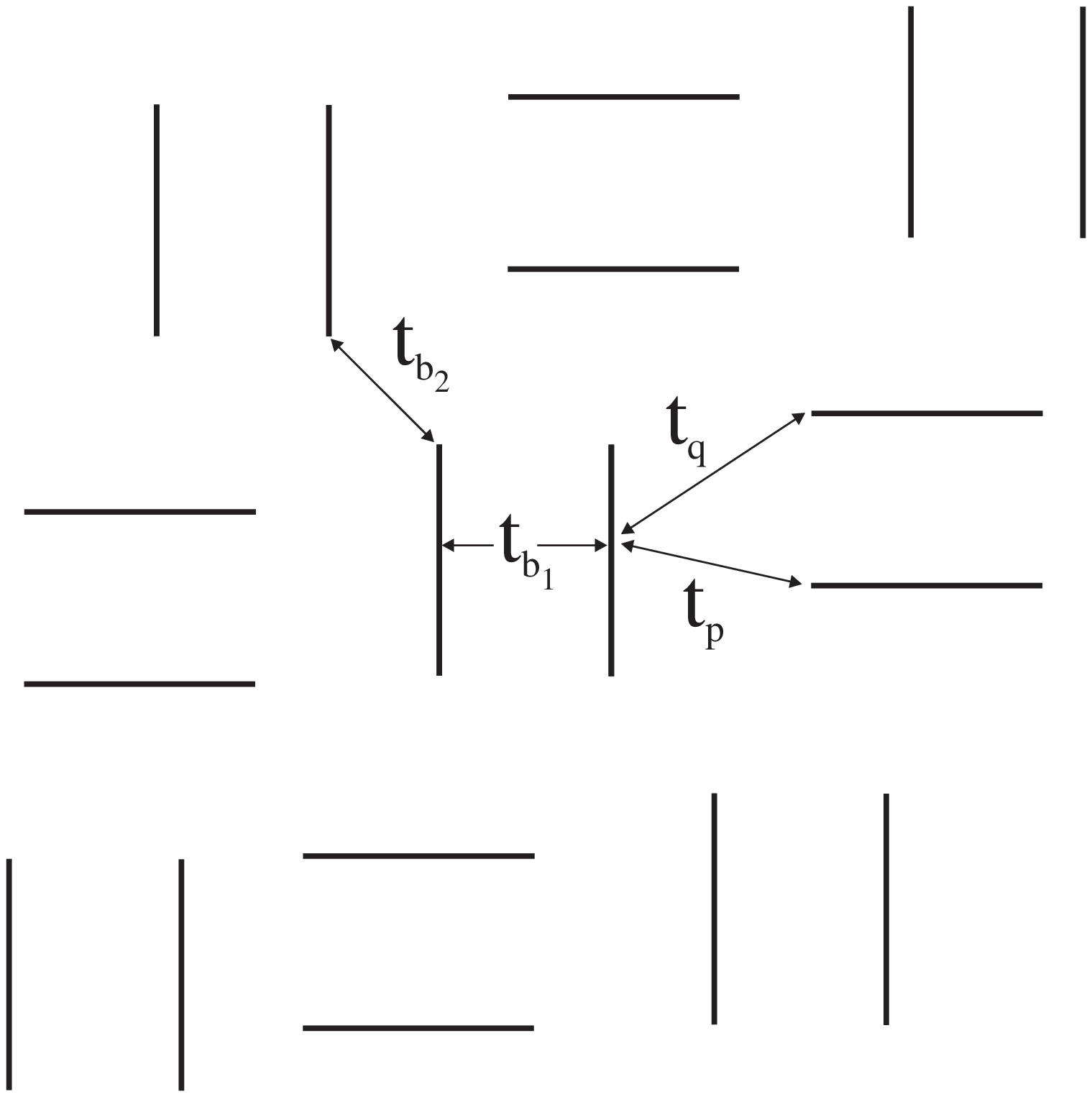}}
\vskip 1.0 cm
\caption{Geometrical arrangement of the BEDT-TTF molecules
in the conducting layers of
the $\kappa$ phase. Each line represents a
BEDT-TTF molecule. The dominant intermolecular hopping integrals
are shown. Some calculated values for these integrals
are given in Table 1.
\label{figx}} \end{figure}
\begin{figure}
\centerline{\epsfxsize=9cm  \epsfbox{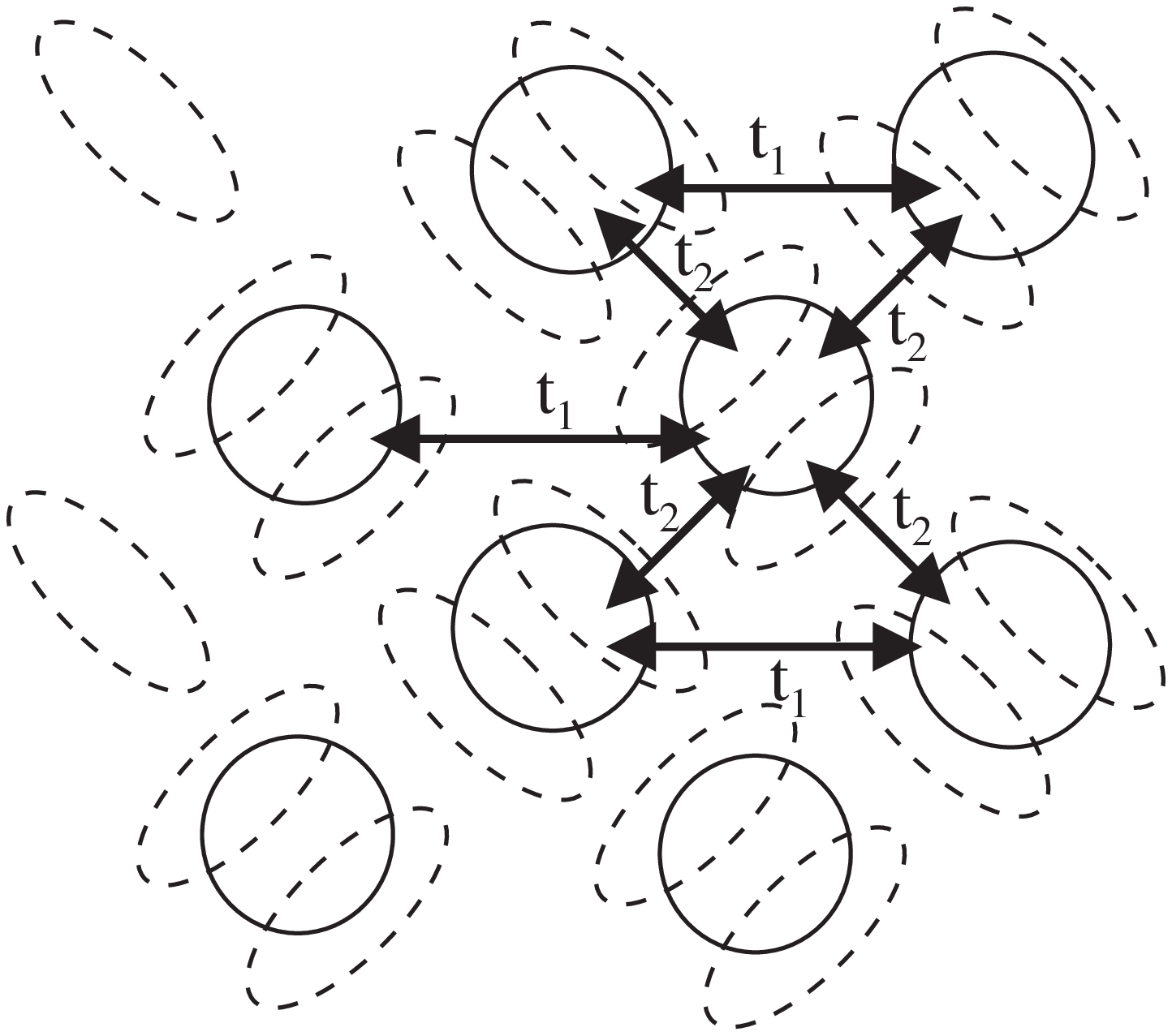}}
\vskip 1.0 cm
\caption{
Triangular lattice for the dimer model of $\kappa-$(BEDT-TTF)$_2$X.
Each lattice site is denoted by a circle and
 represents the anti-bonding orbital on a dimer, a pair of
BEDT-TTF molecules.
Each dashed oval represents a single BEDT-TTF molecule.
There is no hopping in the vertical direction
and there is hopping  along the diagonals.
Competition between the 
two types of hopping results in magnetic frustration
and a Mott-Hubbard metal-insulator transition
at a critical value of the Coulomb repulsion.
\label{fig2}}
\end{figure}
\begin{figure}
\centerline{\epsfxsize=9cm  \epsfbox{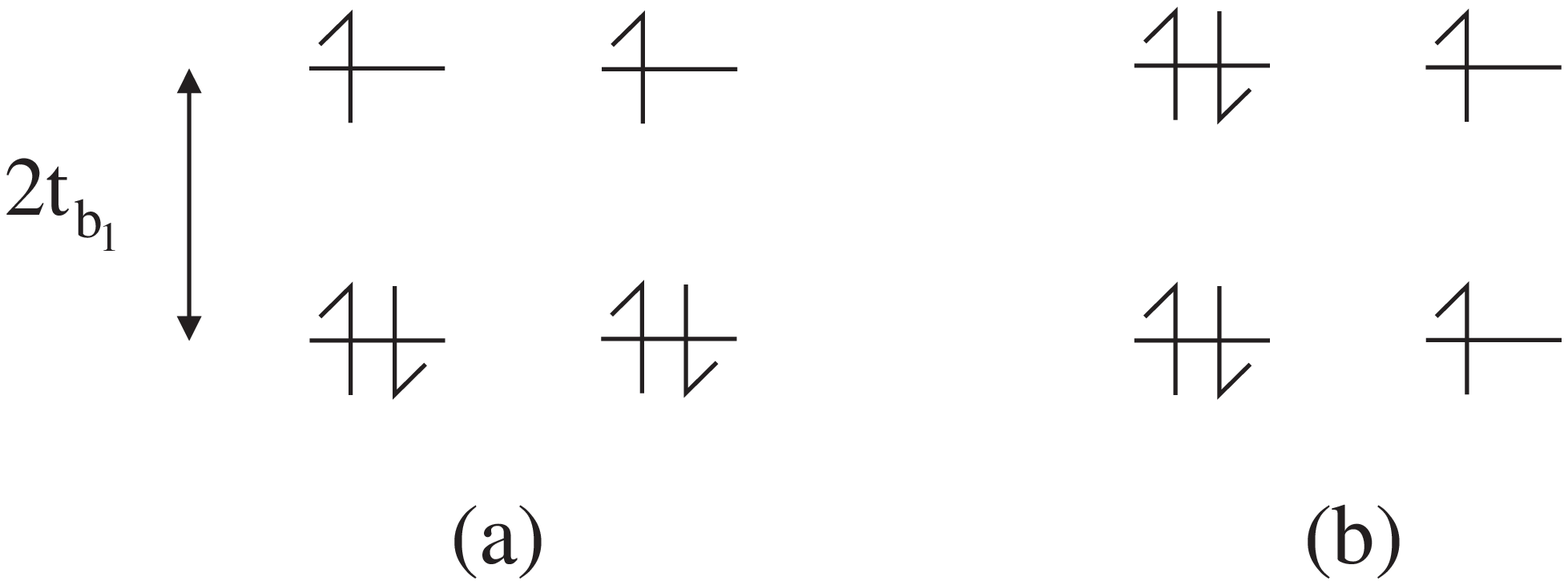}}
\vskip 1 truecm
\caption{
Effective Coulomb repulsion between two electrons
in the anti-bonding orbital of a dimer in the strongly
correlated limit.
(a) The ground state of a pair of three-quarter filled
dimers. If $t_{b_1}$ is the intradimer hopping integral
then $2t_{b_1}$ is the energy splitting of the bonding
and anti-bonding orbital.
(b) If the Coulomb repulsion $U_0$ on a single molecule is
much larger than $t_{b_1}$ then this denotes the configuration
of the lowest energy configuration involving charge transfer
between dimers. The energy difference between (a) and (b) is
$2t_{b_1}$  and so this is the effective Coulomb repulsion
on a dimer for $U_0 \gg t_{b_1}$.
\label{figU}}
\end{figure}
\begin{figure}
\centerline{\epsfxsize=9cm  \epsfbox{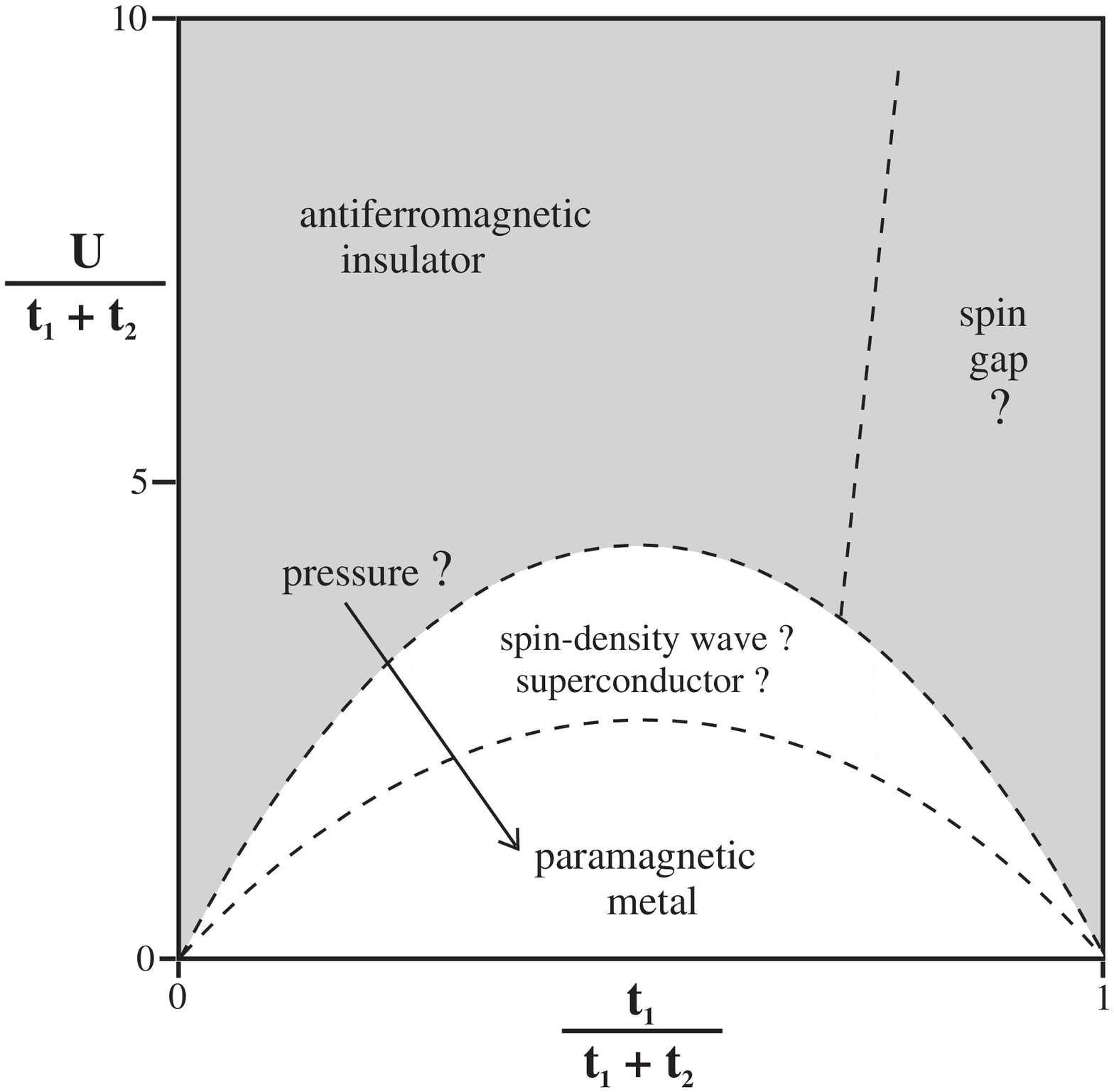}}
\vskip 1 truecm
\caption{
Speculative  phase diagram  for the Hubbard model
on an anisotropic triangular lattice 
(\protect\ref{hubb}).
Shaded and unshaded areas represent insulating and metallic
regions respectively.
 The solid line with an arrow
shows how the ground state of a particular
$\kappa-$(BEDT-TTF)$_2$X crystal might change
as the pressure is increased.
Vertical lines on the left side, center, and right side
correspond to the Hubbard model on a square lattice, 
triangular lattice and decoupled chains, respectively.
\label{fig3}}
\end{figure}
\begin{figure}
\centerline{\epsfxsize=9cm  \epsfbox{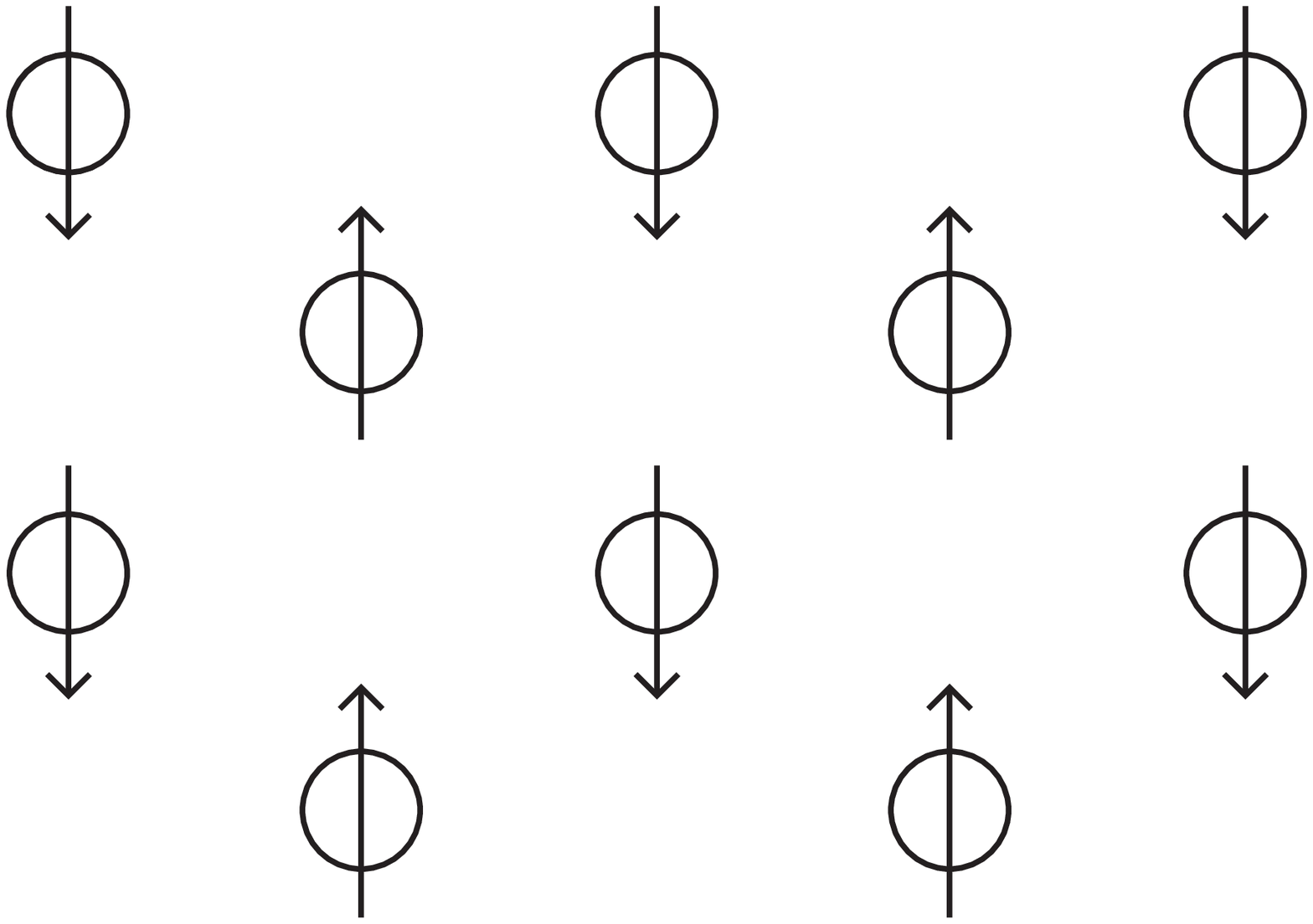}}
\vskip 1 truecm
\caption{
Collinear spin structure of the
antiferromagnetic ground state for the dimer model 
(\protect\ref{hubb})
with $t_1 \ll t_2 \ll U$.
Each circle represents a dimer of BEDT-TTF molecules.
This is the same structure as that proposed for
 deuterated X=Cu[N(CN)$_2$]Br 
based on nmr measurements.\protect\cite{kanoda3}
\label{fig4}} \end{figure}
\begin{figure}
\centerline{\epsfxsize=9cm  \epsfbox{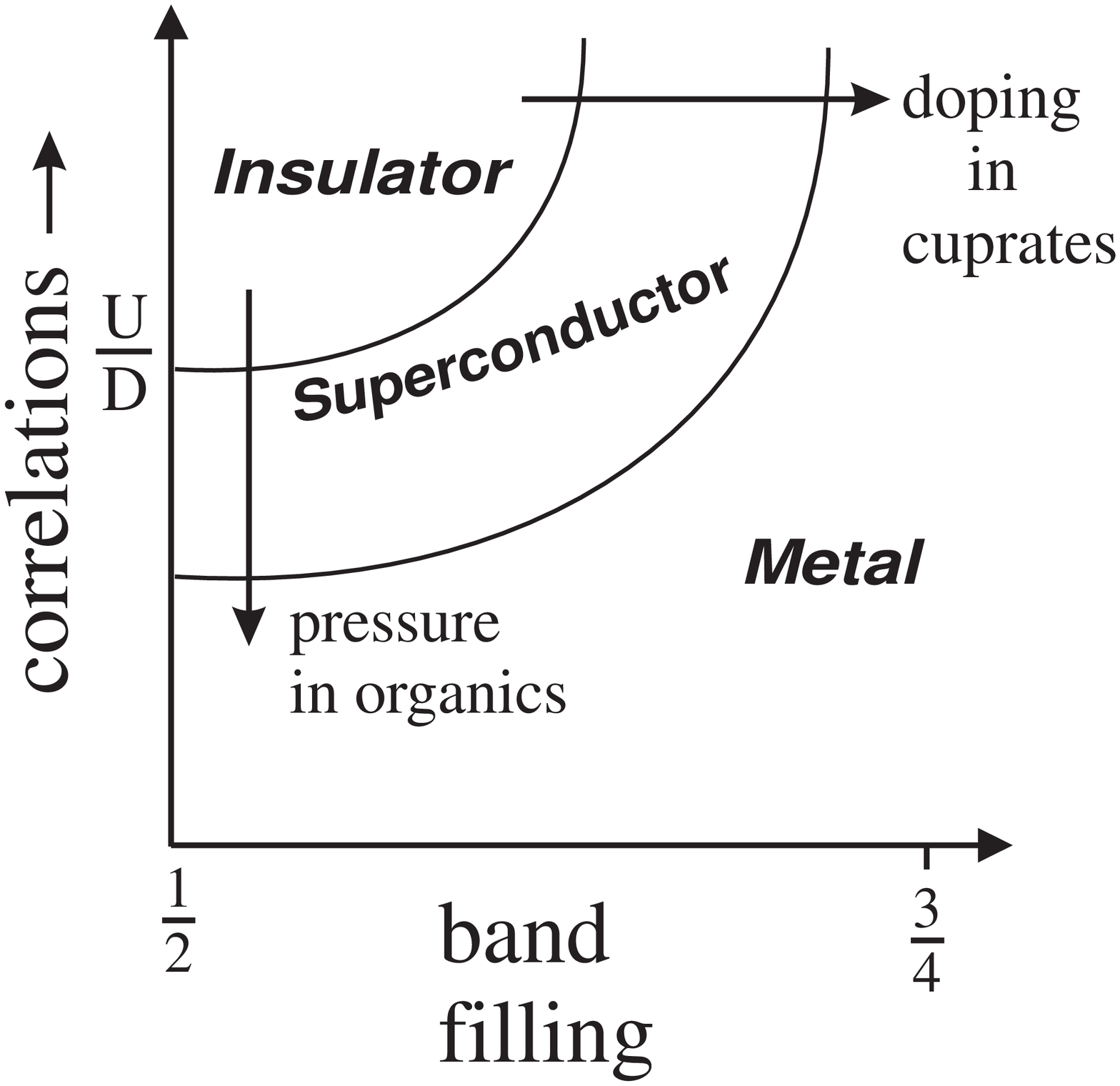}}
\vskip 1 truecm
\caption{
A simple unified picture of the phase diagram of 
the organics and cuprates
at low-temperatures.
The organic superconductors 
$\kappa$-(BEDT-TTF)$_2$X
are always at half-filling and
one tunes through the insulator-superconductor-metal
transition by applying pressure.
In the cuprates 
this transition is passed through by varying the
band filling.
\label{figcup}} \end{figure}

\acknowledgements
This work was supported by the Australian Research Council and
the USA National High Magnetic Field Laboratory
which is supported by NSF Cooperative Agreement
No. DMR-9016241 and the state of Florida.
Helpful discussions with J. S. Brooks, R. J. Bursill,
J. E. Gubernatis, J. Oitmaa, M. J. Rozenberg, J. R. Schrieffer,
R. R. P. Singh, D. Vollhardt, and W. Zheng
 are gratefully acknowledged.
I thank J. Eldridge and Y. Xie for providing the data
shown in Figure \ref{figcond}.
Perez Moses produced the figures.

\vskip 0.5 truecm
                           
{\it Note added in proof.}
Since submission of this article preprints have appeared
which consider the possibility of superconductivity
in the Hubbard model on the anisotropic triangular lattice.
Calculations at the level of the random-phase approximation
(RPA) [M. Votja and E. Dagotto, cond-mat/9807168]
and the fluctuation-exchange approximation
[J. Schmalian, cond-mat/9807042;
H. Kino and H. Kontani, cond-mat/9807147;
H. Kondo and T. Moriya, cond-mat/9807322]
suggest that at the boundary of the antiferromagnetic phase
the model exhibits superconductivity mediated by spin fluctuations.
This result is consistent with the speculative phase
diagram shown in Figure 6 of this article.
As the ratio $t_1/t_2$ increases                       
 the wavevector associated with the antiferromagnetic
spin fluctuations changes from $(\pi,\pi)$ 
and the RPA calculations suggest that
the superconductivity 
changes from d-wave singlet
 to  s-wave triplet in the odd-frequency channel.

\newpage

\newpage

\begin{table}
\caption{
Values of the hopping integrals (in meV) between the
BEDT-TTF molecules in the $\kappa$-(BEDT-TTF)$_2$X
crystal structure for different anions X.
All values were based on quantum chemistry
calculations using the H\"uckel approximation,
except the first line which involved an {\it ab initio} calculation.
All values are for the crystal structure 
at ambient pressure, unless denoted otherwise.
The parameters in the Hubbard model (\protect\ref{hubb})
are given by $t_1 \sim t_{b_2}/2$, $t_2  \sim  (t_p + t_q)/2$,
and $U \sim  2 t_{b_1}$ (See equations (\protect\ref{dimerU})
and (\protect\ref{suppr})).
The fact that $t_1 \sim t_2$ implies
that there is significant magnetic frustration.
Note that for the same anion there is some variation
between the values of the hopping integrals
calculated by different researchers.
} \begin{tabular}{lcrrrccc}
Anion X & $t_{b_1}$ & $t_{b_2}$ & $t_p$ & $t_q$  
 & $t_1/t_2$ & Ref. \\
\tableline
Cu[N(CN)$_2$]Br & 272 & 85 & 130 & 40
 & 0.50& \protect\onlinecite{Fortu} \\
Cu[N(CN)$_2$]Br & 224 & 71 &  94 & 40 
 & 0.53& \protect\onlinecite{Fortu} \\
Cu[N(CN)$_2$]Br & 244 & 92 & 101 & 34
 & 0.68& \protect\onlinecite{komatsu} \\
Cu(CN)$_3$ & 224 & 115 & 80 & 29 
 & 1.10& \protect\onlinecite{komatsu} \\
I$_3$   & 247 &  88 & 119& 33 
 & 0.58& \protect\onlinecite{komatsu} \\
Cu(SCN)$_2$  & 230 & 113 & 99 & 33
 & 0.85& \protect\onlinecite{komatsu} \\
Cu(SCN)$_2$  & 136 &  50 & 41 & 27 
 & 0.74& \protect\onlinecite{rahal} \\
Cu(SCN)$_2$ (7.5 kbar)  & 162 &  54 & 58 & 30
  & 0.61 & \protect\onlinecite{rahal} \\
Cu(SCN)$_2$  & 244 &  26 & 22 & 20
 & 0.62& \protect\onlinecite{campos} \\
Cu(SCN)$_2$ (20 kbar)  & 324 &  45 & 27 & 31
 & 0.88& \protect\onlinecite{campos} \\
\end{tabular}
\label{table1}
\end{table}


\begin{references}
\bibitem{note}
This point was also 
emphasized by Kanoda (Ref. \protect\onlinecite{kanoda3}).
\bibitem{correct}
This effect was overlooked by
Kino and Fukayama.
\end{references}
\end{document}